\documentclass[conference]{IEEEtran}
\IEEEoverridecommandlockouts
\usepackage{cite}
\usepackage{amsmath,amssymb,amsfonts}
\usepackage{algorithmic}
\usepackage{graphicx}
\usepackage{textcomp}
\usepackage{hyperref}
\usepackage{xcolor}
\def\BibTeX{{\rm B\kern-.05em{\sc i\kern-.025em b}\kern-.08em
    T\kern-.1667em\lower.7ex\hbox{E}\kern-.125emX}}

\begin{document}

\title{Blockchains' federation for integrating distributed health data using a patient-centered approach\\
}


\author{\IEEEauthorblockN{Javier Rojo\IEEEauthorrefmark{1}, Juan Hernández \IEEEauthorrefmark{1}, Juan M. Murillo\IEEEauthorrefmark{1} and Jose Garcia-Alonso\IEEEauthorrefmark{1}}

\IEEEauthorblockA{\IEEEauthorrefmark{1}University of Extremadura\\ Caceres, Spain \\
Email:\{javirojo, juanher, juanmamu, jgaralo\}@unex.es} }

\maketitle

\begin{abstract}

Today's world is a globalized and connected one, where people are increasingly moving around and interacting with a greater number of services and devices of all kinds, including those that allow them to monitor their health. However, each company, institution or health system usually store its patients' data in an isolated way. Although this approach can have some benefits related with privacy, security, etc., it also implies that each one of them generates different, incomplete and possibly contradictory views of a patient's health data, losing part of the value that this information could bring to the patient. That is the reason why researchers from all over the world are determined to replace the current institution-centered health systems with new patient-centered ones. In these new systems, all the health information of a patient is integrated into a unique global vision. However, some questions are still unanswered. Specifically, who should store and maintain the information of a given patient and how should this information be made available for other systems. To address this situation, this work proposes a new solution towards making the Personal Health Trajectory of patients available for both, the patients themselves and health institutions. By using the concept of blockchains' federation and web services access to the global vision of a person health can be granted to existing and new solutions. To demonstrate the viability of the proposal, an implementation is provided alongside the obtained results in a potential scenario. 

\end{abstract}


\section{Introduction}


The digitalization of health information systems is a reality that has been present for years. Different health institutions have been digitizing their patients' health records and offering them as Electronic Health Records (EHRs) \cite{kalra2006electronic}. Over the years, and with the undeniable penetration that Internet of Medical Things (IoMT) \cite{Gatouillat2018, rojo2020automating} devices have had among users, these EHRs have been enhanced with information coming from these devices.
In recent years, this has resulted in an evolution from the term EHR to the broader term Personal Health Record (PHR) \cite{Roehrs2017PHR}, which includes any patient's health data; not just those generated and managed in healthcare systems by physicians.


All this digitalization has helped institutions to keep their patients' data organized and presenting a great value for the institution and for the patients themselves. However, due to the globalization of society and the growing number of health solutions, there are more and more institutions and services with which a patient interacts throughout their life \cite{Makitalo2020}.



With the current institution-centered health information systems, this is a problem. The data of a single patient is scattered among the different services of the IoMT devices and health institutions with which they interacts throughout their life. Although this is positive for the institutions, since they have control over the health information they generate, at the same time it implies that there are different representations of a patient's health reality potentially inconsistent between them. In other words, the benefit that these systems generated before for the patient is now mitigated, since the different representations of the same reality make it impossible to have a single global representation of their health: their \textit{Personal Health Trajectory} \cite{Rojo2020}. The creation of health systems that are aware of a patient's health trajectory has been demanded by the fields of medicine and nursing for years \cite{Henly2011}.


To try to solve this problem, an increasing number of studies are trying to give an alternative to the current institution-centered methods of storing patient data. These solutions are patient-centered and try to store each patient's information in a single data structure, with which all institutions and services that treat the patient will interact \cite{Spil2014,kyazze2014design,Prados-Suarez2020,Zhang2015}. For the implementation of this structure, technologies such as blockchain \cite{Roehrs2017,Cichosz2019,Fan2018} are commonly chosen. Blockchain is shown as the most promising technology, due to its distributed nature, as well as its guarantees on data security and privacy \cite{Kassab2019}. However, these proposals are still far from being adopted by the different health institutions, due to the fact that there are unresolved issues, such as the definition of who should keep this data structure alive and who is in charge of granting access to it \cite{Kassab2019}. In addition to these issues, with today's solutions, it is not easy to develop patient-centered health applications because the unique, global health reality is not accessible, limiting innovation and harming patients.


This paper offers a new patient-centered solution that not only provides the advantages of current solutions, but also addresses their weaknesses. For this purpose blockchain technology is used, similarly to existing proposals. However, a more complex architecture has been defined over the concept of one blockchain per patient, which involves the use of blockchains' federation and web services.



Thanks to this proposal, and as demonstrated through the implementation and validation shown in this paper, the integration of patient data is achieved without compromising access to the single global representation of their health by institutions and services, facilitating the development of \textit{Patient Health Trajectory}-aware software and making the transition from actual institution-oriented systems to the new patient-oriented systems closer. In conclusion, the work carried out throughout this paper allows us to offer the \textit{Personal Health Trajectory} of a patient and makes it useful for both patients and the health professionals who care for them.


In order to present the work done, the rest of the document is structured as follows. Section \ref{motivations} shows the motivations of this work and related work. Then, Section \ref{proposal} describes the proposed solution based on blockchains' federation and the use of web services. Section \ref{validation} details the validation performed to evaluate the proposed solution. Section \ref{discussion} brings a brief discussion over the results obtained until now. Finally, in Section \ref{conclusions} the conclusions of this work and some future lines of research are exposed.

\section{Motivations and Related work} \label{motivations}

In the way health systems are conceived today, each health system maintains its patients' data stored in its own information system. When a new patient arrives at a health system, they are registered in the information system and any data produced by the successive interactions they have with that health system will be added. This data is not conditioned by or aware of the data stored by other health systems of their interactions with the same patient. In this way, an isolated representation of each patient in each health system is maintained.

These partial representations of the patient are a problem to be able to offer the best treatment for a patient, since diagnoses are never being offered on the total reality of their health. The problem grows at the same time as the number of services that are generating health data for a patient grows.

To better exemplify the problems this entails, the scenario of Paula is presented. \textit{Paula is a Spanish women who suffers from diabetes. She is also a travel enthusiast. On one of her trips to China, Paula suffers a fainting spell due to hypoglycemia. She is treated in a health center in China, where her blood glucose levels are measured. However, when Paula returns to Spain, she cannot give the data of those measurements to her usual doctor, because they have been registered in the Chinese health system. Something similar happens with her smart glucometer. Only a few of the measurements she takes at home are transcribed to her health record. If Paula has another fainting spell, her doctor cannot access Paula's latest blood glucose measurements nor the ones from the previous episode.}

Providing a way to integrate a patient’s health data and offering it in a way that can track the health of a patient over time can be a solution to the previous problem. This integration can be done by physically integrating the data on the same storage media or by providing a single access point for distributed data that makes it interoperable. In either case, a single, universal vision of the entire health reality of the patient is obtained.

In the first type of solution, there are proposals such as that of Spil et al. \cite{Spil2014} or that of Kyazze et al. \cite{kyazze2014design}, which propose the use of a client-server architecture to store and to consult patients' health data. For this purpose, the first one make use of the Microsoft HealthVault platform, while the second one implement its own architecture called healthTicket. The problem with this type of proposals is that they are very intrusive for health institutions, since they require all institutions to migrate their data to a single repository. In addition to the security and privacy problems that can arise from having all the data in the same repository.

In order to offer less intrusive solutions, researchers such as Zhang et al. propose a client-server architecture \cite{Zhang2015}, but this time using data from distributed repositories managed by different health institutions and services.  Its architecture offers unique access to all this distributed data thanks to the use of a data collection layer and a data management layer, which integrate user data and make them interoperable. Thanks to this, integration is achieved without the need for institutions to migrate their data, although the security problems of the previous proposals remain. There is a single central repository, which makes it a unique point of fault and a point of interest for malicious attacks.

However, if instead of using a client-server architecture to generate the access point, a distributed technology such as blockchain is used, the problems previously mentioned are solved. That is why this technology is used in multiple proposals, such as of Roehrs et al., that propose the use of a blockchain per patient \cite{Roehrs2017}, which stores references to their different PHRs, stored in the institutions and services that generate it. Another proposal of this type is that of Cichosz et al. \cite{Cichosz2019}, although focused on the integration of diabetes data. At the same time, MedBlocks \cite{Fan2018} by Fan et al. proposes a solution similar to the previous one, but also addressing the lack of a standard data management and sharing policy presented by Electronic Medical Records (EMRs), a more primitive representation than the EHRs or PHRs on which the rest of the proposals are focused. 

As can be seen in the previous proposals, the use of blockchain is a step forward in the integration of distributed health data. However, this type of proposal is not free of discussion either. The most important problem is having to manage who owns the structure that maintains this global vision. Normally, the patient holds this role, since data belongs to them. This implies that institutions that want to make use of that global vision of their health will need that the patient gives them permissions to use it. In the case of having to attend a patient in an emergency situation, the patient may not be able to give access to the healthcare professional who is treating them. This limits the effectiveness of the treatment, since the healthcare professional does not have access to data.
At the same time, and also due to the fact that access to the global vision of a patient's health is not easy for institutions and services, the development of software that consumes this global vision is complex in many cases. The existing solutions do not offer in many cases a simple way to develop applications and systems that consume the integrated data of a patient.

The proposal of this paper solves these problems, thanks to the concept of blockchains' federation and the use of web services. A single access point for distributed data is proposed, since the data remains stored in the institutions that generate it, but is organized through a structure that allows access and ensures the interoperability of distributed data.

The blockchains' federation concept is related to proposals such as multi-chain \cite{Kan2018} where the use of a router blockchain to make interoperable a series of lower-level blockchains is proposed. However, these proposals are focused on inter-blockchain communication and increasing the throughput of operations over a single blockchain and provides no improvement for duplications or inconsistencies in the users information. In this case, in the patients information. 

\section{Blockchains' federation and its application to patient-centered health systems} \label{proposal}

The solution proposed in this paper offers unified access to distributed data grouped by patient. For this, an architecture based on blockchains' federation and web services is implemented. This architecture allows the connection with external health applications, which will consume the data through a REST API or a connector. This facilitates the development of health applications that use the \textit{Personal Health Trajectory} of the patients.

As this solution is mainly based on blockchain, it is important that the reader knows at least some previous blockchain concepts, such as block and node. In blockchain, data is stored in an immutable chain of blocks, where blocks are in charge of store the information ---being the chain of blocks the complete set of information---. Each user access this blockchain throughout a node. The nodes are inter-connected between them using a peer-to-peer network. Each of them maintains an complete copy of the blockchain, identical and shared by all the nodes in the network.

\subsection{Blockchains' federation}

\begin{figure}[htb]
\begin{center}
\includegraphics[width=0.45\textwidth]{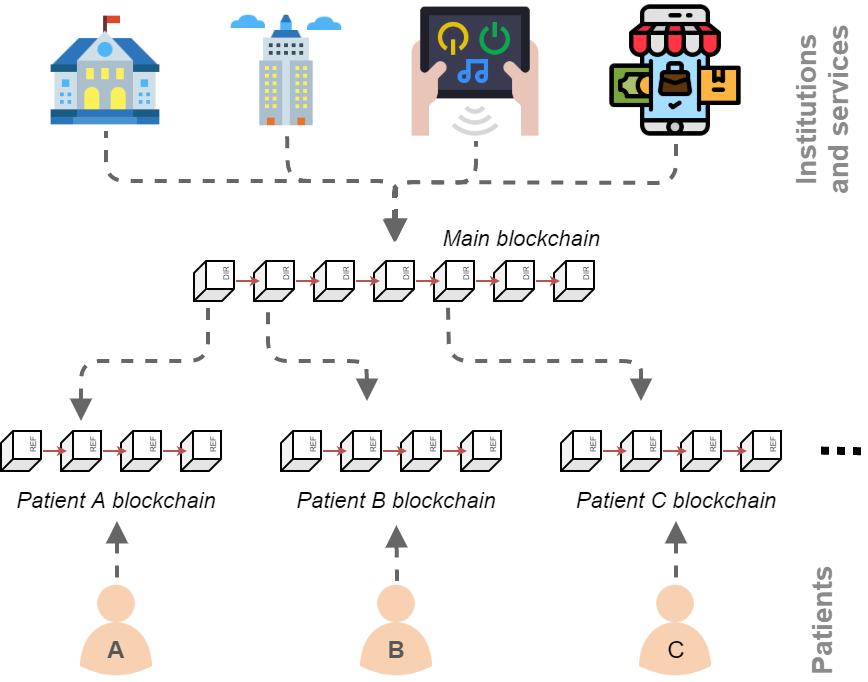}
\caption{Blockchains' federation} 
\label{fig:blockchain_federation}
\end{center}
\end{figure}

The concept of blockchains' federation, as shown in Figure \ref{fig:blockchain_federation}, refers to the interconnection of several lower-level blockchains (patients' blockchains) using another blockchain at the top (main blockchain) which is in charge of nesting them and providing access to all lower-level blockchains.

The patients' blockchains are in charge of saving the information in the system. Each patients's blockchain is self-contained and can be considered and employed as a data structure independent of the federation. The key of the blockchains' federation concept is that each of these blockchains stores the information of one patient. In the case of this proposal, each patient has all their health information, their \textit{Personal Health Trajectory}, organized in a blockchain. Patients can interact directly with their \textit{Personal Health Trajectory} through their blockchain.


To ensure that there are not several partial versions of a patient information in the different institutions and services, each patient must have only one blockchain. In addition, it is necessary to have a mechanism to allow the institutions and services that operated with the data to locate the blockchain of the patient for whom they want to read or write new data. The easiest way to achieve all this is to have a structure that indicates where the blockchain of each existing patient is located. This structure must be shared by all institutions and services. If each of them has its own independent structure to locate the patients' blockchains, the problem is still to ensure that all of them have the same location for the blockchain of one particular patient.

In blockchains' federation, this routing structure is implemented with another blockchain: the main blockchain. Each block in this blockchain stores the location of a patient's blockchain, as well as the information needed to identify the patient to which it belongs. Each institution has one node of the main blockchain, so that any change in the patients or in the location of their blockchains is noted by all institutions and services in the system.

While patients are not aware of the main blockchain and do not interact with it, the institutions and services that produce or consume data for these patients always access the patients' blockchains through it.

\subsection{Architecture}

\begin{figure*}[htb]
\begin{center}
\includegraphics[width=0.70\textwidth]{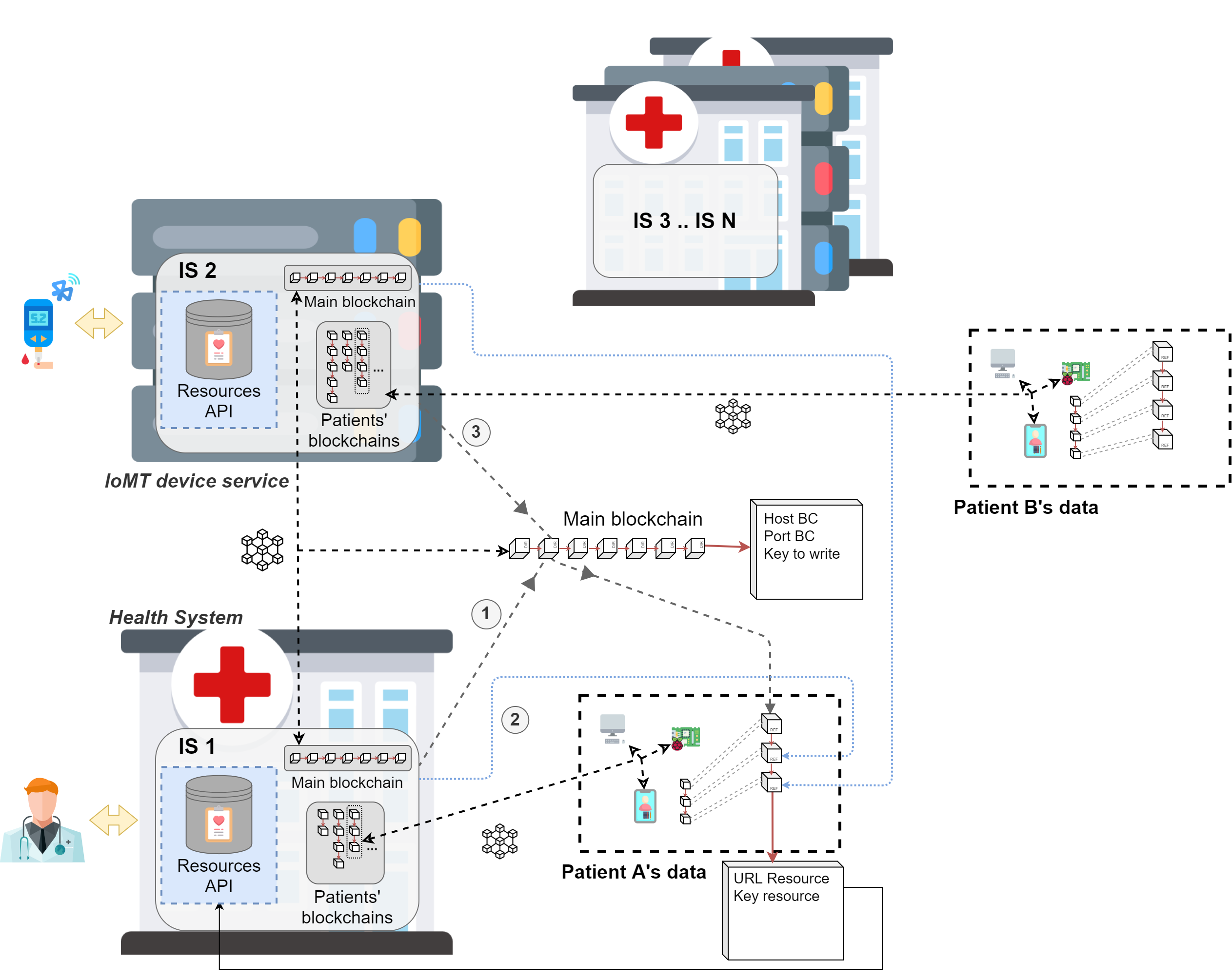}
\caption{Proposed architecture} 
\label{fig:architecture}
\end{center}
\end{figure*}

Using the concept of blockchains' federation presented and web services, the architecture shown in Figure \ref{fig:architecture} is proposed. This architecture addresses the problem of distributed health data integration discussed above, solving the principal limitations of single blockchain proposals while maintaining their advantages. This is mainly thanks to the use of the federation and its main blockchain, which allows the location and access to the patients' data ---stored in the patients' blockchains--- to the institutions and services authorized to use it. In this way, even if a patient is treated urgently and cannot give access to their data, health professionals are able to access them through the main blockchain, as long as their institution belongs to those that integrate their patients' data. 


One of the main components of the architecture is the patients' blockchains. The information they store is private which should only be read and can be created by authorized entities. That's why these blockchains will be permissioned blockchains. 

One of the problems noted in previous proposals about the use of blockchain to integrate patient health data was defining who was in charge of keeping a patient's blockchain alive and accessible. In this proposal, the main health institution to which a patient belongs at a given time is in charge of maintaining at least one node of the patient's blockchain. If the patient changes of main health institution, the new one must deploy a new node of the patient's blockchain. The old one can delete its node at that moment. Additionally, the proposal allows users to have a node of their own blockchain in one of their devices at any time, empowering them with greater control of their \textit{Personal Health Trajectory} and allowing them to extract additional value from it.

As with the other proposals that do not physically integrate the patients' health data, but make them interoperable through a mechanism of references to where the information is actually stored, these patients' blockchains do not directly store the patients' health information. Instead, each block of these blockchains stores a reference to where the health information representing that block is stored. To facilitate this referral system, a Resources APIs is defined. Each healthcare institution and service, in its information system (labeled IS in the figure) deploys a Resources API, as a web service. This Resources API store health data generated at that institution or by IoMT devices using that service, for any patient in the system.
Each time a new health record is added, a URL and key to access it is generated. This URL and key is stored in the blockchain of the patient to which the added health record belongs. This proposed storage system can be easily replaced by any other one, including the current ones in use by most health information systems, as long as they allow to reference the information in a similar way as it is done with the Resources API.


The last component that is involved in this architecture is the main blockchain. In this architecture, the main blockchain is used to locate the blockchain of each patient by the different institutions and services with which they interact. For this purpose, this main blockchain is shared by all of them. Each one of them must have a node of the main blockchain deployed, with which the health professionals of the institutions or the IoMT devices that use the service in question will interact. This blockchain is the component that unifies and ensures that, instead of having several partial views of a patient's health trajectory, there is a single \textit{Personal Health Trajectory} with all the data of that patient. For its implementation, a permissioned blockchain will be used, such as for patients' blockchains. Only entities with enough permissions should be able to access the blockchain to locate the address of the \textit{Personal Health Trajectory} of the different patients.

Figure \ref{fig:architecture} shows the steps that members of the institutions and the services that integrate the data must take to interact with the \textit{Personal Health Trajectory} of a patient. When they want to add or read data from a patient that belongs to the same institution, (1) they employ the main blockchain node of their IS to locate where the patient's blockchain is deployed, and, after locating it, (2) read or write to it. These operation can be similarly performed when the patient does not belong to the same institution. When this happens, instead of creating a new profile or blockchain for the patient with the problems of duplication and inconsistencies already mentioned, they will (3) locate the patient existing blockchain through the main blockchain node of their institution's IS and operate with it.

In the case of being the patients who want to interact with their \textit{Personal Health Trajectory} directly, the process is considerably simplified. Patients are not aware of the existence of the main blockchain, since they will interact directly with their blockchain, through the node maintained by their main institution or through another node they can maintain on one of their devices.

\subsection{Implementation}

The implementation of the proposed architecture depends on the employed blockchain technology, in this case Hyperledger Iroha\footnote{\url{https://github.com/hyperledger/iroha}}. Hyperledger \cite{Dhillon2017} is a project that offers some industrial implementations for permissioned blockchains. Iroha is one of these implementations. Its principal difference with the rest of the Hyperledger's implementations is that it includes a service that allows working with the blockchains without needing a local copy (node). This service, called Torii, allows remote communication with an Iroha's blockchain node. For this purpose, Iroha provides a series of APIs for the main programming languages. In the proposal of this paper, these APIs are employed to create a Python connector and a REST API over it that allow the development of \textit{Personal Health Trajectory}-aware applications and systems without the need of having local copies of the blockchains. This makes it easy to integrate the proposed solution into any kind of device, without the space or computational load limitations of having to host a blockchain node.

All components of the architecture are deployed in the information systems of the institutions and services that integrate their patients' data, except for the Python connector and its REST API. The latter must be deployed along with the healthcare application that employs them. The implementation of each of the components is discussed below and is available in public repositories (see Section \ref{sec:data}). 

\textbf{Resources API.} This component is responsible for storing the health information and make it referable from the blockchain. To do this, it offers a REST API where the information is sent to be stored and where it can be recovered later. When the API receives new information to store, it stores it and establishes a hash identifier and password for that information, which is returned to the sender. That identifier and key are stored in the patient's blockchain, to be employed later, when the information is retrieved. This component supports multi-model storage, so it can store any type of medical data. The format in which it return the data is the one in which it was sent.

\textbf{Patients' blockchains.} Each patient's blockchain is implemented as an Iroha blockchain. This blockchain stores one transaction per block. This transaction represents a change in the patient's health data. If that change is to add, modify or delete information, it is determined by the description of the transaction, so a registry of changes on the patient's information is available for all the involved institutions and services. The resulting registry considering all the changes is the \textit{Personal Health Trajectory} of a patient.

\textbf{Main blockchain.} This blockchain has been also implemented with Iroha. It has at least a transaction for each patient present in an institution. In the description of this transaction it is stored where the patient's blockchain is located, alongside the patient identifier. The process of adding a transaction to the main blockchain must be done by the information system managers, after deploying the patient's blockchain. If any patient changes of principal institution and, therefore, their blockchain changes location, another transaction is generated in this blockchain to store the new address of the patient's blockchain. Therefore, if there are several transactions for the same patient, the address stored in the latest one is always the one considered.

\textbf{Connector.} This component allows software developers to easily develop \textit{Personal Health Trajectory}-aware applications (or health systems), using the proposed architecture as data source. The connector can be integrated into any software application, similar to a database connector. It allows users to add and retrieve information from the \textit{Personal Health Trajectory} of a given patient. In order to use it, it is needed to connect to one of the nodes of the main blockchain or to a blockchain's node of a patient.

\textbf{Connector REST API.} For those developers who prefer to deploy the connector as a web service to which all their applications connect, a REST API is provided that offers the connector methods via HTTP requests.

\subsection{Enabling development of Personal Health Trajectory-aware software}

\begin{figure}[htb]
\begin{center}
\includegraphics[width=0.48 \textwidth]{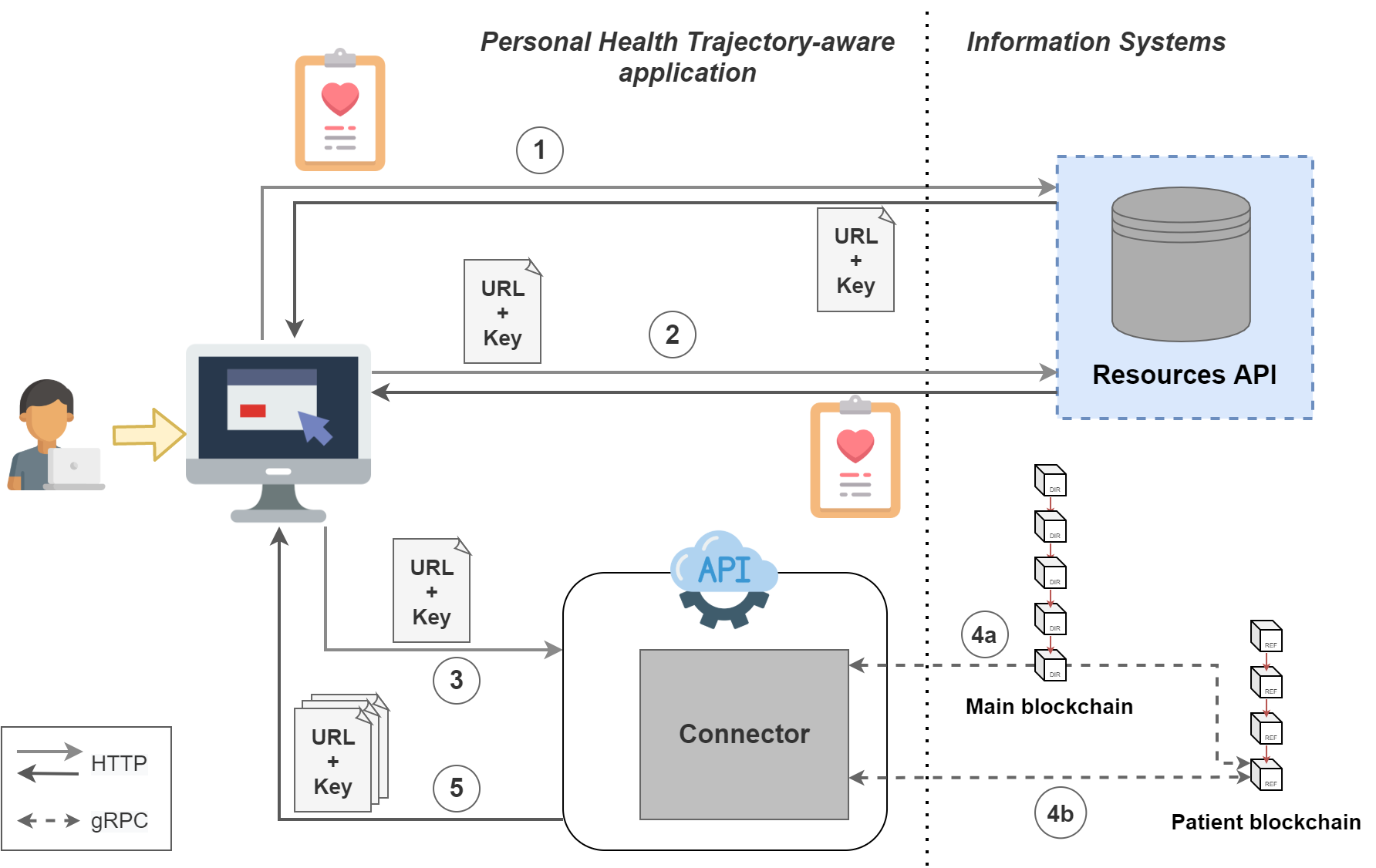}
\caption{Connections scheme for a Personal Health Trajectory-aware application} 
\label{fig:phtaware_apps}
\end{center}
\end{figure}

Although it has been mentioned that the proposed architecture offers a connector to develop \textit{Personal Health Trajectory}-aware software, it has not been discussed yet which are the steps or considerations to develop this type of software. Figure \ref{fig:phtaware_apps} shows a basic diagram of the connections that any \textit{Personal Health Trajectory}-aware application must have with the components of the proposed architecture, deployed alongside the application or in the information systems of the institutions and services that integrate their patients' data.

\textit{Personal Health Trajectory}-aware applications interact with the architecture through two components. 

On the one hand, an application of this type interacts with the storage system in charge of storing the evidences or records in the different information systems. By default the Resources APIs, if no information system has chosen a different storage method. It interacts with this component to save the records and generate a URL and key to access them (\textit{1}) and to retrieve the records with the URL and key that identifies each one of them (\textit{2}). The records can be retrieved from any Resources API, the URL will indicate exactly where it is stored, but the process of adding an evidence can only be performed on the Resources API of the information system to which the application has granted access. For example, in the case of a doctor's application, the records are added in the Resources API of the institution to which the doctor belongs. In this way, the records remain stored in the institution or service that generates them.

On the other hand, the application also interacts with the Connector. This interaction can be done by directly including the Connector within the application or by deploying it as a web service to be used by the application, as in Figure \ref{fig:phtaware_apps}. The Connector allows storing the references generated by the interactions with the Resources API (\textit{1}) in the blockchain of the patient to which the record belongs (\textit{3}). The patient's identifier must be attached to this information, in case the Connector has been configured to use the main blockchain (\textit{4a}) in order to interact with the \textit{Personal Health Trajectory} of any patient. In this case, the Connector must have been previously configured with the address and private key of the main blockchain node with which it will operate, normally that of the information systems of the institution or service to which the application belongs. In the case of being configured to work only with the \textit{Personal Health Trajectory} of a patient, it is not necessary to send the patient's identifier to which the record to be added belongs, but it is necessary to previously configure the Connector with the address and the private key of the patient's blockchain node with which it will operate. This way, the Connector will connect directly to this blockchain (\textit{4b}), without going through the main blockchain.

The Connector also allows the retrieval of the set of references to records that make up the \textit{Personal Health Trajectory} of a patient (\textit{5}). To do this, it connects again with the patient's blockchain node, directly (\textit{4b}) or through the main blockchain (\textit{4a}), in the same way as explained for the process of adding new references.

If an entity develop several \textit{Personal Health Trajectory}-aware applications or systems that interact with the architecture across the same node of the main or patient's blockchain, the more interesting option is to deploy the Connector as a web service and employ this Connector REST API with all the applications.

\section{Validation} \label{validation}

In order to evaluate the technical suitability of the proposal ---discarding social and organizational issues---, an initial validation of the proposal has been performed. For this, the proposed architecture has been deployed in the information systems of a series of simulated healthcare institutions. A web application has also been developed using Angular and Bootstrap. It uses the Connector REST API to allow doctors to operate with the \textit{Personal Health Trajectory} of the patients of these organizations.

The developed application allows doctors from different institutions to access the data and add new evidences to patients of any of the deployed institution. In this application, the evidences stored are always documents, of any format, since this is enough to check the viability of the proposal. The results obtained with documents can be extrapolated to any type of information ---stored as JSON, YAML,...---, due to the fact that information stored in blockchains ---references--- and the Resources APIs's web services are independent of the evidences format. The implementation of the application is also available in a public repository (see Section \ref{sec:data}). 


\begin{figure}[h]
\begin{center}
\includegraphics[width=0.48\textwidth]{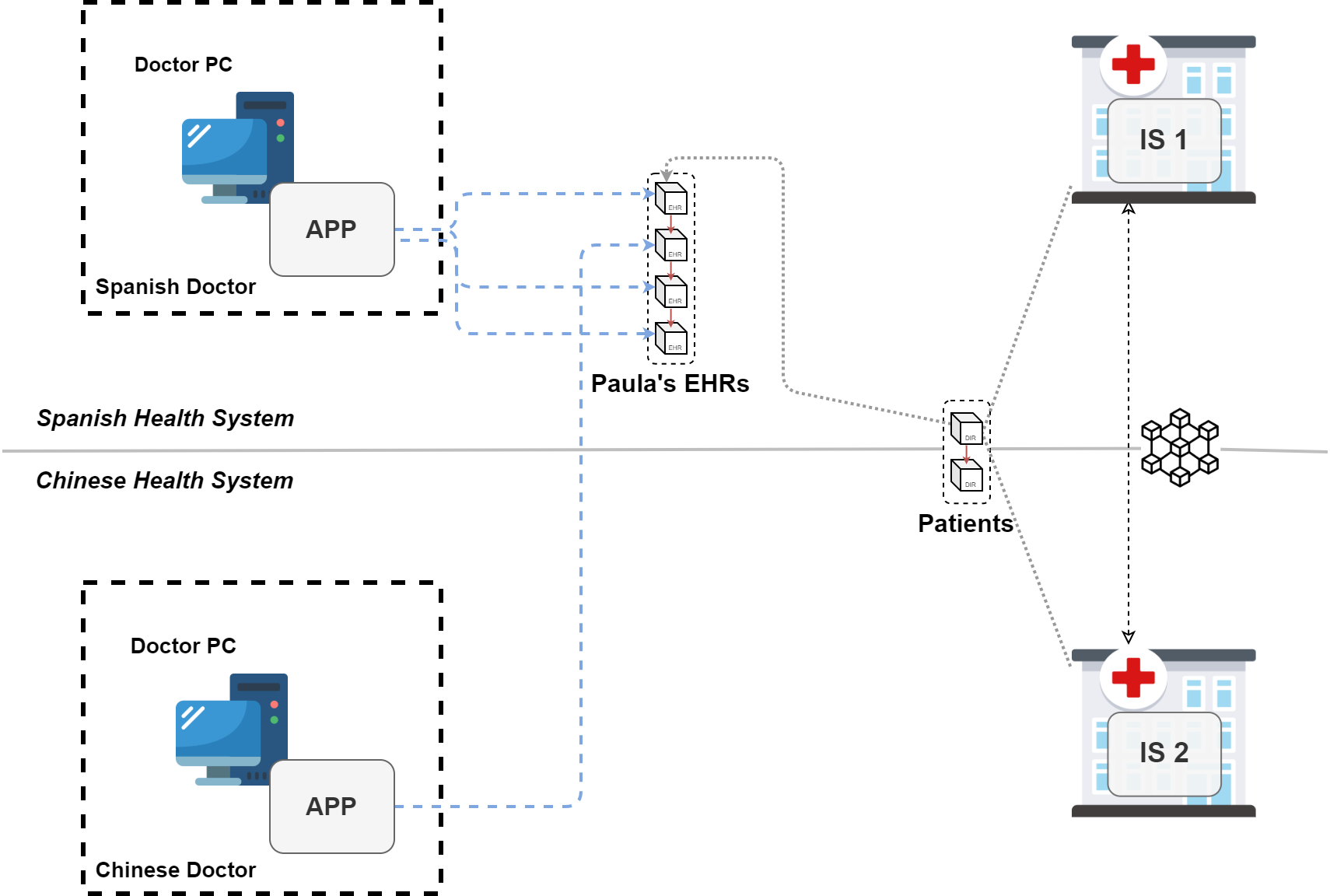}
\caption{Scenario where the proposal is employed to integrate information of a patient since two institutions.} 
\label{fig:case_of_use}
\end{center}
\end{figure}

The scenario ---Figure \ref{fig:case_of_use}---  on which this validation is focused is the one of Paula, discussed in Section \ref{motivations}. Therefore, the proposed architecture is used to simulate the integration of patient data from the Spanish and Chinese health systems. In this way, when Paula is treated by the Chinese health system doctor, the evidence of this intervention will be added to her \textit{Personal Health Trajectory}. When she returns to Spain, her usual doctor can check the data from the intervention that Paula suffered in China and continue working on them.

The main hypothesis that leads to this validation is to check that the integration of the data is being done correctly and is accessible by the different institutions without patient interaction. In addition, it is checked which are the delays in the doctor's use experience due to the fact that all the mentioned architecture is behind generating the integration of the data. These delays are measured in terms of the time needed to add new data for a patient and to recover it from their \textit{Personal Health Trajectory}.

The deployment of the architecture has been carried out in the following environment:

\begin{itemize}
    \item \textbf{Information Systems:} Two EC2 instances from Amazon Web Services with Ubuntu 18.04 LTS, 1 virtual core at 2.5GHz and 1GB of memory.
    \item \textbf{Doctors computer:} a laptop with Windows 10, 16GB of RAM and an Intel Core i7-8550U processor at 1.8GHz base frequency and 4.0GHz turbo frequency. It is equipped with NVMe SSD technology for storage.
\end{itemize}

\subsection{Results}

A series of numerical and non-numerical results have been obtained from the validation of the proposal of this paper.

The first of these results has been to confirm the main hypothesis that guided this validation. Paula's health data have been correctly integrated and the doctors of both institutions have been able to work on a unique, global and complete version of her health reality.

The second part of the results includes the evaluation of the times that the architecture has taken to perform the different operations that have been carried out. The value of these times has been measured using 20 execution of the operations. The average value for each of them can be seen in Table \ref{tab:validation}.

\begin{table}[h]
\begin{center}
\caption{Validation results}
\begin{tabular}{| c | c |}
\hline
Time measurement & Value (s) \\ \hline
Locate patient's blockchain & 0.079 \\
Save evidence's resource & 0.21 \\
Add evidence to patient's blockchain & 4.784 \\
Retrieve evidence's resource & 0.209 \\
Recover evidences from patient's blockchain & 0.128 \\ \hline
\end{tabular}
\label{tab:validation}
\end{center}
\end{table}

From the table above, it can be seen that the most expensive operation is to add a change in the \textit{Personal Health Trajectory} of the patient (\textit{Add evidence to patient's blockchain}), due to the fact that a new block must be created in their blockchain for recording the change in their health information. 
Because the main blockchain is not altered in the process of adding changes to patient data, this cost is independent of the number of institutions and services that maintain a main blockchain node. Once a patient is added to the system, this main blockchain is only used to search where their blockchain is located. However, the number of nodes in the patient's blockchain can already influence the execution times for adding new information, due to the need of consensus between nodes. The data collected in Table \ref{tab:validation} shows the execution time when there is only one node of the patient's blockchain. 

As for the times to locate a patient in the main blockchain (\textit{Locate patient's blockchain}) and to retrieve the list of changes stored in their blockchain (\textit{Recover evidences from patient's blockchain}), it has been considered that they do not negatively impact the use experience of the doctor's application, being its values of the order of a few milliseconds. This operation is not affected by the number of nodes of the main blockchain or the patient's blockchain, because in the reading operations it is not necessary a consensus between the different nodes of a blockchain.

The times for recovering and adding files through the Resources API (\textit{Save evidence's resource} and \textit{Retrieve evidence's resource}) depend only on the size of the file and the bandwidth of the internet connection, so they are not discussed.

\section{Discussion} \label{discussion}

After validating the proposal and commenting on the results, this section discusses if the proposal makes a really useful contribution to the world of distributed health data integration. We assume that the results obtained were those expected and that the main hypothesis of the validation has been fulfilled.

The first thing to indicate is that the main contribution of the proposed architecture is not the use of a blockchain to store references to all the health data of a patient. This is already done by other proposals, where the suitability of this technology is already sufficiently proven. On the opposite, it should be considered that the main contribution that this proposal makes to this area of research is the use of the blockchains' federation to promote easier access to integrated patient information by the institutions and to keep alive the structure that holds the data of each patient integrated. In this sense, this proposal has solved the problems discussed in the introduction ---without renouncing the benefits already offered by current proposals--- by which the use of blockchain as a method to develop patient-centered health systems still generates some doubts. The institutions and services will continue to keep the data they generate in their systems, but now they have a structure that allows them to operate with a global vision of patient health, and not with the fragmented visions with which they operated until now. This is a benefit both for them and for the patients.

In addition, the proposal facilitates the transition to patient-centered health systems by offering a Connector that simplifies the development of health applications and systems that consume the \textit{Personal Health Trajectory} data of patients. Thanks to this Connector and a well-defined procedure for the development of \textit{Personal Health Trajectory}-aware applications, the proposed architecture is really useful for health software developers.
In this way, it encourages health software developers to start using this proposal as data source for their developments.



Furthermore, with the proposal presented in this paper, the suitability of blockchain to develop patient-centered health systems has only been reaffirmed. In the first place, and as already indicated in previous works, blockchain has shown itself to be adequate due to its qualities of  decentralization, sharing and openness guaranteeing traceability and security in the data as well. All this obtained by means of the ledger abstraction which model the shared unique record of the transactions between different systems. In second place, because the different advantages that have been indicated throughout the proposal and that have been possible thanks to blockchain. For example, federating the different patients' blockchains with another blockchain allows all institutions and services to be aware of the changes in patients' blockchain locations at the same time and to work with the same blockchain for one patient. The fact that the structure that maintains the \textit{Personal Health Trajectory} of a patient is another blockchain, empowers the patient with their data, allowing them to have a node of it on their devices and ensuring that no one malicious can generate changes in their data. Therefore, with all this, the suitability of blockchain for this proposal is more than clear.

In any case, the proposal of this paper is not free of debate either. The proposed architecture has tried to be as less intrusive as possible for the institutions and services, providing Resources APIs and Connector, even allowing them to maintain their storage system if it allows referencing the information from the blockchain. However, institutions and services, even if they do not replace their information systems, need to include a series of components in them.

\section{Future works and conclusions} \label{conclusions}

The proposal of this paper offers a method to build patient-centered health systems, by integrating patient data, which is distributed among the different institutions and services that generate it, in a unique and global vision: its \textit{Personal Health Trajectory}. Thus, institutions and the patients themselves can operate on a complete version of patients' health information.

To do this, the concept of blockchains' federation is defined and applied in a proposed architecture. This architecture defines a series of components that software developers can use to create applications or systems that consume the architecture's data. These types of applications ---\textit{Personal Health Trajectory}-aware applications--- and the procedure for connecting them to the proposed architecture has been defined.

As future steps in the proposal, a more complex validation of the architecture is expected, with better performance measures, scalability tests ---with more blocks and more nodes, both for patients and main blockchain---, stress tests and formal comparison with other type of solutions, like a traditional database or single blockchain proposals.

\section{Data Availability} \label{sec:data}

The software developed is available in Zenodo:
\begin{itemize}
    \item \textbf{IS components}: \href{https://doi.org/10.5281/zenodo.4588545}{10.5281/zenodo.4588545}
    \item \textbf{Connector (and its API)}: \href{https://doi.org/10.5281/zenodo.4588534}{10.5281/zenodo.4588534}
    \item \textbf{Doctor app}: \href{https://doi.org/10.5281/zenodo.4588555}{10.5281/zenodo.4588555}
\end{itemize}

\section*{Acknowledgment}

This work was supported by the project 0499\_4IE\_PLUS\_4\_E funded by the Interreg V-A España-Portugal 2014-2020 program, by the project RTI2018-094591-B-I00 (MCIU/AEI/FEDER, UE), by the FPU19/03965 grant, by the Department of Economy and Infrastructure of the Government of Extremadura (GR18112, IB18030), and by the European Regional Development Fund.

\bibliographystyle{IEEEtran}
\bibliography{main}

\end{document}